\documentstyle[11pt]{article}
\newlength{\bredde}
\def\slash#1{\settowidth{\bredde}{$#1$}\ifmmode\,\raisebox{.15ex}{/}
\hspace*{-\bredde} #1\else$\,\raisebox{.15ex}{/}\hspace*{-\bredde}
#1$\fi}
\newcommand{\mat}{\left ( \begin{array}{cc}}
\newcommand{\emat}{\end{array} \right )}
\newcommand{\matt}{\left ( \begin{array}{ccc}}
\newcommand{\ematt}{\end{array} \right )}
\newcommand{\matf}{\left ( \begin{array}{cccc}}
\newcommand{\ematf}{\end{array} \right )}
\newcommand{\vect}{\left ( \begin{array}{c}}
\newcommand{\evect}{\end{array} \right )}

\newcommand{\be}{\begin{eqnarray}}
\newcommand{\ee}{\end{eqnarray}}
\newcommand{\beq}{\begin{equation}}
\newcommand{\eeq}{\end{equation}}
\newcommand{\ba}{\begin{array}{ccc}}
\newcommand{\ea}{\end{array}}

\newcommand{\noi}{\vspace{12pt}\noindent}
\newcommand{\lG}{\raise.3ex\hbox{$\stackrel{\leftarrow}{G}$}}
\newcommand{\lU}{\raise.3ex\hbox{$\stackrel{\leftarrow}{U}$}}
\newcommand{\lP}{\raise.3ex\hbox{$\stackrel{\leftarrow}{{\cal P}}$}}
\newcommand{\leta}{\raise.3ex\hbox{$\stackrel{\leftarrow}{\eta}$}}
\newcommand{\lOmega}{\raise.3ex\hbox{$\stackrel{\leftarrow}{\Omega}$}}
\newcommand{\ldr}{\raise.3ex\hbox{$\stackrel{\leftarrow}{\delta^r}$}}

\def\beqn{\begin{eqnarray}}
\def\eeqn{\end{eqnarray}}

\def\gtwid{\raise.3ex\hbox{$>$\kern-.75em\lower1ex\hbox{$\sim$}}}
\def\ltwid{\raise.3ex\hbox{$<$\kern-.75em\lower1ex\hbox{$\sim$}}}

\begin{document}

\title{
{\vspace{-1.5cm} \normalsize
\hfill \parbox{40mm}{CERN-TH/2001-077}}\\
{\vspace{-0.35cm} \normalsize
\hfill \parbox{40mm}{Edinburgh 2001-01}}\\
{\vspace{-0.35cm} \normalsize
\hfill \parbox{40mm}{NBI-HE-01-04}}\\[25mm]
\Large{\bf The C-Theorem and Chiral Symmetry Breaking\\
in Asymptotically Free Vectorlike Gauge Theories}}
\vspace{1.5cm}

\author{~\\{\sc R.D. Ball}\footnote{On leave from:
Department of Physics and Astronomy, University of Edinburgh, 
Mayfield Road, Edinburgh EH9 3JZ, Scotland.} \\~\\and\\~\\
{\sc P.H. Damgaard}\footnote{On leave from:
The Niels Bohr Institute, Blegdamsvej 17, DK-2100 Copenhagen,
Denmark.}\\
\\\\
Theory Division\\CERN\\CH-1211 Geneva 23\\Switzerland
}
\maketitle
\vspace{1.5cm}
\begin{abstract}
We confront Cardy's suggested $c$-function for four-dimensional field
theories with the spontaneous breaking of chiral symmetries 
in asymptotically free vectorlike gauge theories with 
fermions transforming according to different representations under the 
gauge group. Assuming that the infrared limit
of the $c$-function is determined by the dimension of the associated
Goldstone manifold, we find that this $c$-function always decreases
between the ultraviolet and infrared fixed points.
\end{abstract}
\vfill

\thispagestyle{empty}
\newpage


The $C$-theorem in two dimensions \cite{Z} demonstrates the
irreversibility
of renormalization group flows by construction of a $c$-function which
is proven to decrease montonically along the flow. As such it is an
important
ingredient in our understanding of nonperturbative field theories.
For four dimensional theories Cardy suggested a particular
$c$-function \cite{Cardy} which he likewise conjectured to 
decrease along renormalization group trajectories. There is some 
controversy over whether this $C$-theorem for four-dimensional 
field theories has been proven \cite{FL} or not \cite{OS} (see also
refs.
\cite{JO,CFL,Cap}). Studies of supersymmetric
theories with duality symmetries \cite{B,Dan} have found that 
in all cases Cardy's $c$-function does decrease along the RG flow. 
For non-supersymmetric gauge theories, so far the only nonperturbative
case that has been considered is that of QCD: Cardy in his original paper
\cite{Cardy} found that indeed the proposed $c$-function did decrease 
towards the infrared when the number of flavors $N_f$ of Dirac fermions 
was in a range compatible with asymptotic freedom, and chiral symmetry 
is spontaneously broken. 

In this letter we will extend this analysis to consider the $c$-function
for all asymptotically free vectorlike gauge theories with Dirac 
fermions. Our main assumption
will be that Cardy's conjectured form of the $c$-function may be used 
all the way towards the infrared, where chiral 
symmetry may be broken spontaneously, and where at the fixed point the only 
massless excitations are those of the associated Goldstone manifold. 
Comparison of the value of the $c$-function in the ultraviolet and 
the infrared will then allow us to test the conjectured $C$-theorem
(or conversely put constraints on the allowed pattern of chiral 
symmetry breaking if we find that for a particular pattern the 
theorem is violated). We will consider systematically all 
possible simple compact gauge groups and all possible irreducible 
representations $r$ of
$N_f$ Dirac fermions coupled vectorially to the gauge fields. 

The number $N_f$ and the representation
$r$ that the fermions carry will only be constrained by the demand that
the theory must be asymptotically free.
As we exclude from the beginning the presence of fundamental scalars,
the one-loop $\beta$-function takes the form 
\beq
\beta(g) = -\frac{g^3}{16\pi^2}\left[\frac{11}{3}\ell({\cal G}) -
\frac{4}{3}
\ell(r)N_f\right] + \ldots ~, \label{betaf}
\eeq
where $\ell({\cal G})$ and $\ell(r)$ are the indices of the
representations
carried by the gauge bosons and the fermions, respectively. These
indices 
$\ell$ are defined by
\beq
\ell \delta_{ab} = 2{\rm Tr}(T_aT_b)
\eeq
where $\{T_a\}$ are the generators of the group in the particular 
representation: they are closely related to the Casimirs $C$
\beq
C\delta_{ij} = (T^aT^a)_{ij}
\eeq
(see, $e.g.$, ref. \cite{Slansky} for a review). Because
we wish to maintain asymptotic freedom we impose the constraint 
\beq\label{af}
N_f ~<~ \frac{11}{4}\frac{\ell({\cal G})}{\ell(r)} \label{AF}
\eeq
on the number of flavors of fermions. It may well be that this 
bound is too weak, and that both confinement
and chiral symmetry breaking are lost for smaller values of $N_f$
(as, for example, when there is a perturbative infrared fixed point in
the gauge coupling). But this bound must at least always be satisfied.
We take all $N_f$ fermions to be strictly massless.

Cardy's proposal for a $c$-function in four dimensions is based on the
Euler
term in the trace of the energy-momentum tensor. In the natural
normalization
where $c$ is unity for one massless scalar degree of freedom
\cite{Cardy}, it
takes the form \cite{CD} 
\beq
c ~=~ N_0 + 11N_{1/2} + 62 N_1 \label{cdef}
\eeq
at the fixed points. Here
$N_0$ counts the number of massless real scalars, $N_{1/2}$ the 
number of massless Dirac fermions, and $N_1$ 
the number of massless vector bosons. Let next $d({\cal G})$ denote the
dimension of the gauge group ${\cal G}$, and $d(r)$ the dimension of
the irreducible representation $r$. The idea is now quite simple:
we compare the value of $c$ in the ultraviolet, where for an 
asymptotically free gauge theory we have the ``fundamental'' 
gauge and fermion degrees of freedom, and thus
\beq
c_{UV} ~=~ 11d(r)N_f + 62 d({\cal G}) ~,
\eeq
with the value in the infrared, where the only massless degrees
of freedom are assumed to be those of the Goldstone bosons which 
arise from chiral symmetry breaking. If $d(G/H)$ is the 
dimension of the Goldstone manifold $G/H$, this implies
\beq
c_{IR} ~=~ d(G/H) ~.
\eeq 
If the $C$-theorem holds, we should find that $c_{UV} \geq c_{IR}$.

To determine the number of Goldstone bosons we need to know the 
pattern of chiral symmetry breaking $G \to H$ in strongly coupled 
(confining) vectorlike gauge theories, when $N_f$ (Dirac) fermions
transform 
according to an irreducible representation $r$ of the gauge 
group ${\cal G}$. There are three generic classes of breaking 
to consider:\footnote{For an excellent exposition of these ideas,
see $e.g.$ ref. \cite{Peskin}.} 
\begin{itemize}
\item The fermion representation $r$ is pseudo-real: chiral symmetries
are enhanced from $SU(N_f)\times SU(N_f)$ to $SU(2N_f)$, and the
expected
symmetry breaking pattern is $SU(2N_f) \to Sp(2N_f)$.
\item The fermion representation $r$ is complex: the expected 
symmetry breaking
pattern is $SU(N_f)\times SU(N_f) \to SU(N_f)$.
\item The fermion representation $r$ is real: chiral symmetries are again
enhanced to $SU(2N_f)$, and the expected symmetry breaking pattern is
$SU(2N_f) \to SO(2N_f)$.
\end{itemize}

\noindent Because of a connection between the Dyson classification in
Random Matrix Theory and the three symmetric spaces $G/H$ discussed
above
\cite{Zirn,V}, it has become customary to label these three symmetry
breaking patterns by their Dyson indices $\beta = 1, 2$ and 4,
respectively. 
Of course, there can be exceptions to these scenarios, when for instance
the
gauge theories contain fundamental scalars, or additional
symmetries (such as supersymmetry). We will not consider such cases in
this paper.

It is not easy to prove that these symmetry breaking patterns actually 
do occur dynamically. Two specific cases have been
proven by Coleman and Witten \cite{CW}: the symmetry 
breaking pattern of fermions in the
defining representations of gauge groups $SU(N_c)$ or $SO(N_c)$
in the large-$N_c$ limit (cases $\beta = 2$ and $\beta=4$, respectively). 
The one class that is being left out in the Coleman-Witten paper is the
one for which the fermions transform as a pseudo-real representation
of the gauge group. This can be achieved by considering gauge groups
$Sp(2N_c)$. Taking the large-$N_c$ limit, we easily extend the
arguments of ref. \cite{CW}, and thus prove that indeed the chiral
$SU(2N_f)$ symmetry in that case breaks down to $Sp(2N_f)$, as expected
from the above classification.
Anomaly matching conditions \cite{thooft} can also be used to constrain 
the form of vacuum condensates \cite{VKS}. 

For our purposes, however, it is sufficient to note that the 
three generic symmetry breaking patterns each assume maximal 
breaking of chiral symmetry consistent with the preservation of 
maximal flavor symmetry. The spontaneous breaking of flavor 
symmetries in vectorlike gauge theories is prohibited by the 
Vafa-Witten theorem~\cite{VW}. It follows that  
each pattern gives an upper bound on the number of Goldstone bosons in 
the broken theory, and thus an upper bound on the $c$-function in the 
infrared. Thus if the $C$-theorem is satisfied with maximal breaking,
it will be satisfied in general.

The counting is thus very straightforward in the infrared. For the three
different classes of maximal breaking we have:
\be
c_{IR}&=\left\{ \begin{array}{ll} \mbox{$d(SU(2N_f)/Sp(2N_f))$} 
&\\ 
\mbox{$d(SU(N_f)\times SU(N_f)/SU(N_f))$} &\\ 
\mbox{$d(SU(2N_f)/SO(2N_f))$} &\end{array}\right.
=\left\{ \begin{array}{ll} \mbox{$N_f(2N_f-1)-1$} &\mbox{for
$\beta=1$},\\ 
\mbox{$N_f^2-1$} &\mbox{for $\beta=2$},\\ 
\mbox{$N_f(2N_f+1)-1$} &\mbox{for $\beta=4$}.\end{array}\right.
\ee
The value taken by Cardy's $c$-function in the
infrared is thus bounded above by a function of $N_f$ and the 
value of $\beta$. Note that in the special case $N_f=1$, for the symmetry
breaking classes of $\beta=1,2$, the Goldstone manifold is zero-dimensional.
There are then no Goldstone bosons at all, and the only breaking of chiral
symmetry is the explicit one due to the anomaly. The theory then has
a mass gap in the infrared, and it is consistent to set $c_{IR}=0$ in
those cases. For the $\beta=4$ class the Goldstone manifold is non-trivial
even for $N_f=1$, however, consistent with the fact that this case is 
equivalent to having two Majorana fermions.

It is easy to show that
\beq
N_f^2-1~\leq~N_f(2N_f-1)-1~<~N_f(2N_f+1)-1, 
\mbox{\qquad for all \qquad $N_f\geq 1$,}
\eeq 
so that for a given $N_f$ the tightest $C$-theorem constraints are 
found when the fermions are in real representations $r$, i.e. when 
$\beta=4$. We can use this observation 
to prove the following lemma:\\

{\it Lemma:} if $c_{UV}\geq c_{IR}$ for fermions in a real  
representation $r_0$ of ${\cal G}$ with dimension $d(r_0)$ and 
index $\ell(r_0)$, it will also hold for all other representations 
$r$ of ${\cal G}$ with dimension $d(r)\geq d(r_0)$ and index 
$\ell(r)\geq\ell(r_0)$. 

The proof is straightforward: $d(r)\geq d(r_0)$ means that $c_{UV}$ is 
greater for $r$ than for $r_0$, while $\ell(r)\geq\ell(r_0)$ means that 
the condition (\ref{af}) on $N_f$ from asymptotic freedom is more
relaxed.
If $r$ is not real but pseudo-real or complex, $c_{IR}$ is reduced 
for a given $N_f$. \\

Given that the adjoint representation is always real, confirming the 
$C$-theorem for the adjoint is thus sufficient to confirm it for all 
representations except for a finite number of low dimension. In general 
these must then be considered on a case by case basis.
We will be systematic, and go through all the simple compact Lie 
groups in the Cartan classification, beginning with the orthogonal and 
symplectic groups, through the unitary groups to the exceptional
groups.\\

\noindent{\bf $SO(N_c)$:} Although from the group-theoretical 
perspective the $SO(N_c)$ groups are very
different depending on whether $N_c$ is even or odd, here we may
consider
all the orthogonal groups together. We take $N_c>6$, since $SO(2)$ is 
abelian, and $SO(3) \sim SU(2)$, $SO(5) \sim Sp(4)$ and $SO(6)\sim
SU(4)$
will all be treated below. The dimension of $SO(N_c)$ is
$d({\cal G}) = N_c(N_c-1)/2$, and the index of the adjoint
representation is $\ell(SO(N_c)) = N_c-2$, so the condition of
asymptotic freedom becomes
\beq
N_f ~<~ \frac{11}{4}\frac{N_c-2}{\ell(r)} \label{AFSO}
\eeq
First consider the defining representations of $SO(N_c)$: these are 
all real, so $\beta=4$, and furthermore have $\ell(r)=1$ and 
$d(r) = N_c$. In the ultraviolet we thus have
\beq
c_{UV} ~=~ 11N_cN_f + 31N_c(N_c-1),
\eeq
while in the infrared 
\beq
c_{IR} ~=~ N_f(2N_f+1)-1 ~.
\eeq
The condition $c_{UV} \geq c_{IR}$ thus becomes
\beq
N_f ~\leq~ \frac{11}{4}(N_c-2) + \frac{21}{4} +
\frac{3}{4}\sqrt{41N_c^2-30N_c+1}
\eeq
which is automatically satisfied since the condition of asymptotic
freedom (\ref{AFSO}) in this case gives
\beq
N_f ~<~ \frac{11}{4}(N_c-2) ~.
\eeq

For all other representations $r$ of $SO(N_c)$ we can use the lemma,
since 
$\ell(r)\geq 1$ and $d(r) \geq N_c$. So the $c$-function decreases
between the ultraviolet and the infrared for all representations
of all $SO(N_c)$ groups.\\ 

{\bf $Sp(2N_c)$:} The adjoint representations of $Sp(2N_c)$ have 
indices $\ell({\cal G}) = 2(N_c+1)$,
and $d({\cal G}) = N_c(2N_c+1)$ for these gauge groups.
The condition of asymptotic freedom (\ref{AF}) thus becomes
\beq
N_f ~<~ \frac{11}{2}\frac{N_c+1}{\ell(r)} ~.
\eeq
All representations of $Sp(2N_c)$ are either real or pseudo-real, and
the
fundamental representations of $Sp(2N_c)$ are always pseudo-real. 
Thus for the fundamental representations
\beq
c_{UV} ~=~ 22N_cN_f + 62N_c(2N_c+1) ~,
\eeq
while
\beq
c_{IR} ~=~ N_f(2N_f-1) - 1 ~.
\eeq
The condition $c_{UV} \geq c_{IR}$ then translates into
\beq
N_f ~\leq~ \frac{11}{2}N_c + \frac{1}{4} + \frac{3}{4}
\sqrt{164N_c^2+60N_c+1} ~, \label{spcond}
\eeq
while the asymptotic freedom condition for the fundamental
representation 
requires
\beq
N_f ~<~ \frac{11}{2}(N_c+1) ~.
\eeq
One sees that the inequality (\ref{spcond}) is satisfied for all $N_c$. 
Because we again here have
$\ell(r)=1$ for the fundamental representation, and all other pseudoreal 
representations have both larger $\ell(r)$ and larger dimensions $d(r)$,
we conclude using an argument similar to that used for the lemma that 
the $c$-function decreases from the ultraviolet to
the infrared for all pseudo-real representations of $Sp(2N_c)$.

We next turn to the real representations of $Sp(2N_c)$. 
The smallest real representation is not the adjoint, but
rather a representation with dimension $d(r)=N_c(2N_c-1)-1$
and index $\ell(r)=2(N_c-1)$. For this represention 
the condition $c_{UV} \geq c_{IR}$ becomes
\beq
N_f ~\leq~ \frac{1}{4}\left[11N_c(2N_c-1)-10 + 
\sqrt{(11N_c(2N_c-1)-10)^2 +8(62N_c(2N_c+1)+1)}\right] ~,
\label{spcondpr}
\eeq
which is easily satisfied for all $N_c\geq 2$ since the 
asymptotic freedom condition is now
\beq
N_f ~<~ \frac{11}{4}\frac{N_c+1}{N_c-1} ~.
\eeq
Since all other real representations have both larger 
$\ell(r)$ and larger dimensions $d(r)$, and combining this with our result
above for all pseudo-real representations of $Sp(2N_c)$,
we conclude that the $C$-theorem 
is satisfied for all representations of $Sp(2N_c)$. \\

{\bf $SU(N_c)$:} We begin with gauge groups $SU(N_c), N_c \geq 3$. 
The fundamental representations are
all complex, they have the common index $\ell(r)=1$, and the symmetry
breaking
class is the one of $\beta=2$. The condition $c_{UV} \geq c_{IR}$ gives
\beq
N_f ~\leq~ \frac{11}{2}N_c + \sqrt{121N_c^2/4 + 62N_c^2 - 61} ~.
\label{QCD}
\eeq
Because the condition (\ref{AF}) in this case reads $N_f < 11N_c/2$
the condition (\ref{QCD}) is, as observed already by Cardy \cite{Cardy}, 
automatically satisfied. Since all other complex representations of 
$SU(N_c)$ have both dimension and index greater than that of the
fundamental,
it follows from an argument similar to that used for the lemma that
the $C$-theorem is satisfied for all complex representations of 
$SU(N_c)$. 

But the groups $SU(N_c)$ have, in
general, both real and pseudo-real representations too. Because $d(r)$
is not a monotonic function of $\ell(r)$, it is not possible to select
the ``lowest'' real or pseudo-real representations, check those, and
then conclude that all other real or pseudo-real representations will
yield a decreasing $c$-function too. We can, however, consider all the
adjoint representations of $SU(N_c)$. They are real, and we therefore
have
\beq
c_{UV} ~=~ (N_c^2-1)(11N_f+62) ~,
\eeq
while
\beq
c_{IR} ~=~ N_f(2N_f+1)-1 ~,
\eeq
which leads to
\beq
N_f ~\leq~ \frac{1}{4}\left[11N_c^2-12+
\sqrt{(11N_c^2-12)^2+8(62N_c^2-61)}\right] ~,
\eeq
a bound which is very much above the requirement of $N_f<11/4$ from
asymptotic
freedom. So for all adjoint representations of $SU(N_c)$ we have
$c_{UV} > c_{IR}$. By the lemma, the same will be true of all 
representations with dimension and index greater than those of the
adjoint.

For the rest, it seems that the only solution is
to consider the remaining real and pseudo-real representations of 
the $SU(N_c)$ groups on a case-by-case basis. We have done this 
for all non-Abelian gauge groups of rank less than 9, based on 
the tables of ref. \cite{Patera}.
For $SU(3)$, $SU(5)$, $SU(7)$ $SU(8)$ and $SU(9)$ it turns out that 
there are no such representations which also satisfy the asymptotic 
freedom constraint and have smaller dimension and index than the
adjoint.
The group $SU(4)$ has no pseudo-real representations,
and three real representations with $\ell(r) < 22$, as required from
asymptotic freedom. The one with lowest index ($\ell(r)=2$) has
dimension
6, which gives
\beq
c_{UV} ~=~ 66N_f + 930 ~~,~~~~~~ c_{IR} ~=~ N_f(2N_f+1)-1 ~,
\eeq
and hence a bound of $N_f \geq (65+ 3\sqrt{1297})/4$, which is much above
the bound $N_f < 11$ from asymptotic freedom. Because the two other
real representations have both larger indices and larger dimensions,
they are even further away from the bound imposed by asymptotic freedom.
$SU(6)$ has one relevant representation of index $\ell(r)=6$, which is 
pseudo-real and of
dimension 20. One easily checks that $c_{UV} > c_{IR}$. 

Finally we treat the special case of the gauge group $SU(2)$. Here the
condition for asymptotic freedom (\ref{AF}) becomes $\ell(r)N_f<11$. The
fundamental representation is pseudo-real, and the symmetry breaking
class is
the one of $\beta=1$. We thus have 
\beq
c_{UV} ~=~ 22N_f + 186 ~~,~~~~~~ c_{IR} ~=~ N_f(2N_f-1)-1 ~,
\eeq
and hence the condition $N_f \leq 17$, which well encompasses the bound
from
asymptotic freedom of $N_f < 11$. By a similar argument to that used for 
the lemma, this implies that $c_{UV} > c_{IR}$ for all other pseudo-real 
representations of $SU(2)$. It remains to check only the 
real representations, those of integer
isospin $j$, which correspond to the symmetry breaking class of
$\beta=4$. The index of an isospin-$j$ representation in $SU(2)$ is
\beq
\ell(r) ~=~ \frac{2}{3}j(j+1)(2j+1) ~.
\eeq
Because of the bound from asymptotic freedom of $\ell(r) < 11$ this 
means that we need only consider the adjoint ($j=1$) representation for
which
$\ell(r) = 4$. Here,
\beq
c_{UV} ~=~ 33N_f + 186 ~~,~~~~~~ c_{IR} ~=~ N_f(2N_f+1) -1 ~,
\eeq
which leads to $N_f < 8 + (3/2)\sqrt{70}$, which again is well above the
bound $N_f < 4$ from asymptotic freedom. We conclude that
$c_{UV} > c_{IR}$ for all representations of $SU(2)$.\\

\noindent
{\em The Exceptional Groups:}

For $E_6$ the adjoint index is $\ell({\cal G})=24$, which gives as
condition
for asymptotic freedom $\ell(r)N_f < 66$. There are then only two
relevant
representations: the fundamental of index $\ell(r)=6$ and dimension 27,
and the adjoint which has dimension 78. The fundamental representation
is complex. In both cases we find that $c_{UV} > c_{IR}$.

In $E_7$ the adjoint index is $\ell({\cal G})=36$, and the condition of 
asymptotic freedom is thus $\ell(r)N_f < 99$. There are then two 
relevant representations, the fundamental of $\ell(r)=12$ and dimension
56, 
and the adjoint which has dimension 36. The fundamental representation 
is pseudo-real. In both cases we again find $c_{UV} > c_{IR}$. 

The exceptional group $E_8$ is special in that the
fundamental representation coincides with the adjoint. By the lemma 
this is the only relevant representation: it has  $d({\cal G})=248$ 
and $\ell({\cal G})=60$, so the condition of asymptotic freedom 
is $N_f < 11/4$.  It is easy to check that $c_{UV} > c_{IR}$. 

For gauge group $F_4$
the adjoint index is $\ell({\cal G})=18$, which gives $\ell(r)N_f <
99/2$ from
the condition of asymptotic freedom. There are two relevant
representations,
the fundamental of index $\ell(r)=6$ and dimension 26, which is real, 
and the adjoint of dimension 52. By the lemma, only the fundamental need
be checked, and we get $c_{UV} > c_{IR}$. 

Finally, the group $G_2$ has adjoint index $\ell({\cal G})=2$, which
leads to
$\ell(r)N_f < 22$ from the demand of asymptotic freedom. There are then
three relevant representations, all real: the fundamental of 
$\ell(r)=2$ and dimension 7, the adjoint of dimension 14, 
and an $\ell(r)=18$ representation of dimension 27. Again by the lemma 
it is only necessary to check the fundamental, where $c_{UV} > c_{IR}$.\\ 

In conclusion, we have undertaken a systematic 
examination of Cardy's proposed 
$c$-function in the context of asymptotically free vectorlike gauge 
theories. Our main assumption has been that we
can follow the renormalization group flow from the ultraviolet gauge 
theory degrees of
freedom nonperturbatively to an infrared fixed point where the only 
massless excitations are those of the Goldstone bosons. We also assume 
that the pattern of symmetry breaking is consistent with the Vafa-Witten
theorem, and that there are no exotic phenomena such as 
exactly massless bound-state fermions in the infrared theory. 
We have found that the $c$-function decreases along the RG trajectories 
in all cases considered: for all gauge groups, with fermions in any 
representation of the gauge group, and for all allowed patterns of symmetry
breaking. Turning this around, we find that
a $C$-theorem would impose no new constraints on the pattern of 
spontaneous chiral symmetry breaking in asymptotically free 
vectorlike gauge theories.

It may be of interest to apply the constraints imposed by the
$C$-theorem
on theories in which the above assumptions no longer hold. In
particular, 
perhaps we can learn more about the possible
symmetry breaking patterns in chiral (but anomaly-free) gauge theories
in this way: for chiral theories the Vafa-Witten theorem no longer 
holds and the pattern of spontaneous symmetry breaking may be more
involved. 
It may also be of interest to consider constraints on chiral symmetry 
breaking imposed by other criteria \cite{ACS}. 

\noi\noindent
{\bf Acknowledgments}\\
We would like to thank Graham Shore for comments. The work of 
R.D.B. and P.H.D. was supported in part by EU TMR grant nos.
ERBFMRXCT97-0122 and ERBFMRXCT98-0194 respectively.


\end{document}